\documentclass[a4paper, reprint, twoside]{revtex4-1}

\usepackage[english]{babel}
\usepackage[utf8]{inputenc}
\usepackage[labelfont=up]{subcaption}
\usepackage[x11names]{xcolor}
\usepackage{float}
\usepackage{mathtools}
\usepackage[top=1.in, bottom=1.in, left=0.5in, right=0.5in]{geometry}
\usepackage{amsthm, amssymb}
\usepackage{graphicx}
\usepackage{lipsum}
\usepackage{bm}
\usepackage[pdftex, pdftitle={Article}, pdfauthor={Author}]{hyperref}
\usepackage{wasysym}
\usepackage{enumitem}
\hypersetup{colorlinks=true, citecolor=black, linkcolor=black, urlcolor=blue}

\DeclareMathOperator{\me}{e}

\DeclarePairedDelimiter\ton{(}{)}
\DeclarePairedDelimiter\qua{[}{]}

\DeclarePairedDelimiter\mean{\langle}{\rangle}

\usepackage{color} 

\newcommand{\blue}[1]{{\color{blue} #1}}

\begin{document}

\title{Quantifying the complexity and similarity of chess openings using online chess community data}
\author{Giordano De Marzo$^{1, 2, 3, 4}$}
\author{Vito D. P. Servedio$^{4}$}
    
    \affiliation{$^1$Centro Ricerche Enrico Fermi, Piazza del Viminale, 1, I-00184 Rome, Italy.\\
    $^2$Dipartimento di Fisica Universit\`a ``Sapienza”, P.le A. Moro, 2, I-00185 Rome, Italy.\\
    $^3$Sapienza School for Advanced Studies, ``Sapienza'', P.le A. Moro, 2, I-00185 Rome, Italy.\\
    $^4$Complexity Science Hub Vienna, Josefstaedter Strasse 39, 1080, Vienna, Austria.
    }

\date{\today} 


\begin{abstract}
Hundreds of years after its creation, the game of chess is still widely played worldwide. Opening Theory is one of the pillars of chess and requires years of study to be mastered. Here we exploit the ``wisdom of the crowd'' in an online chess platform to answer questions that, traditionally, only chess experts could tackle. We first define the relatedness network of chess openings that quantifies how similar two openings are to play. In this network, we spot communities of nodes corresponding to the most common opening choices and their mutual relationships, information which is hard to obtain from the existing classification of openings. Moreover, we use the relatedness network to forecast the future openings players will start to play and we back-test these predictions, obtaining performances considerably higher than those of a random predictor. Finally, we use the Economic Fitness and Complexity algorithm to measure how difficult to play openings are and how skilled in openings players are. This study not only gives a new perspective on chess analysis but also opens the possibility of suggesting personalized opening recommendations using complex network theory.
%
\end{abstract}

\maketitle

\section*{Introduction}
\begin{figure*}[t]
			\centering
			\includegraphics[width=0.9\textwidth]{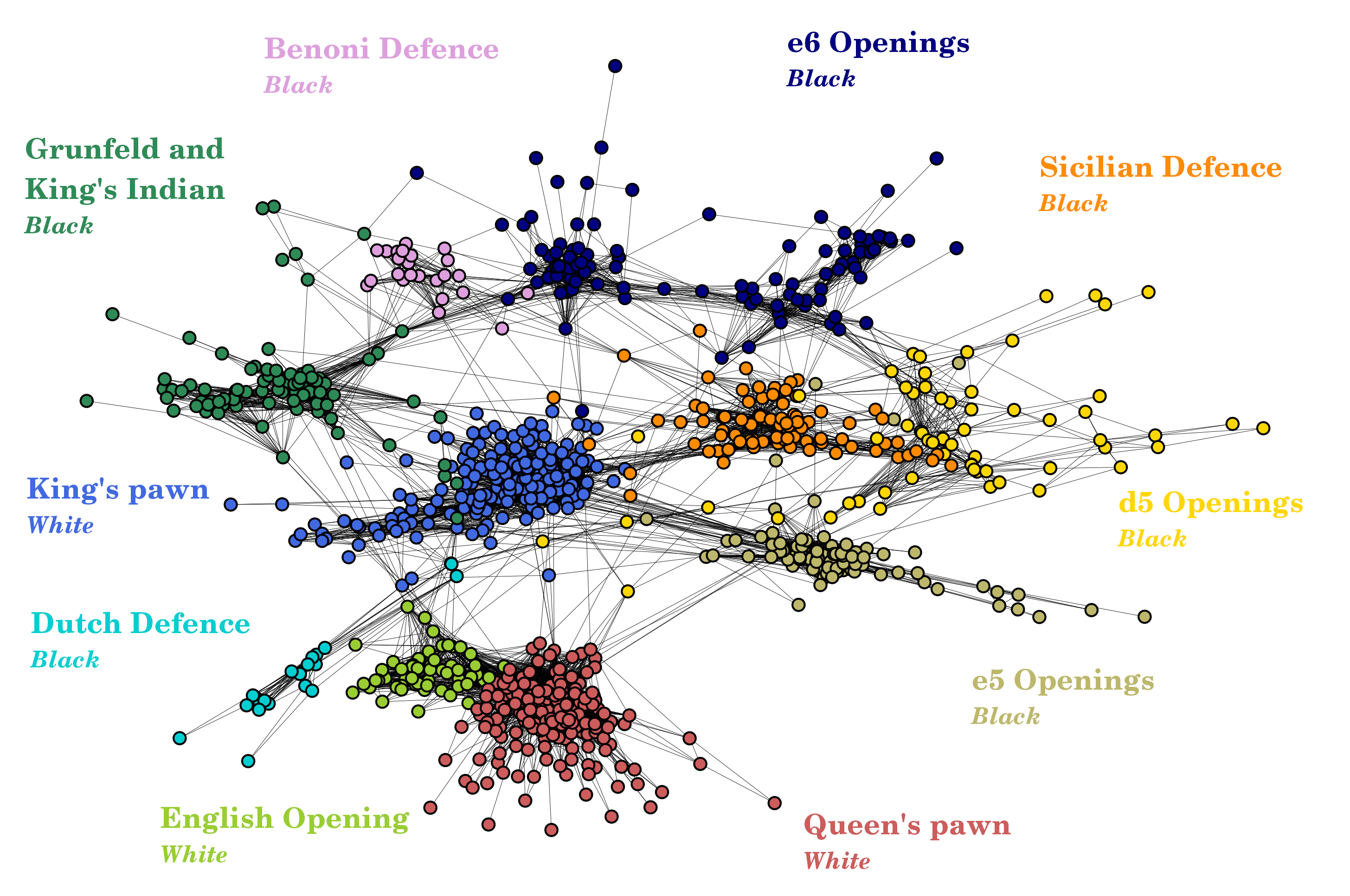}
			\caption{\textbf{Network of openings.} Relatedness network of openings obtained by projecting and validating the player-opening bipartite network. Openings close on this network are similar to play since they often appear together in players' opening repertoires. Using the Leiden community detection algorithm, we identified ten clusters that compose the network. These clusters are represented in the network using different colors, each corresponding to a different opening choice.}     
			\label{fig:network_leiden}
		\end{figure*}
Since its creation in the 6th century, chess has fascinated an uncountable number of people and nowadays, 1500 years after its birth, it counts more than 600 million regular players (\url{https://www.un.org/en/observances/world-chess-day}). Chess, which is considered by many to be among the noblest intellectual arts, has strongly influenced human history. For instance, chess has been one of the main arenas where the Soviet Union and the United States fought for intellectual supremacy, as exemplified by the 1972 world championship match between Bobby Fisher and Boris Spassky. Iconic is also the match between world champion Garry Kasparov and IBM supercomputer Deep Blue, won by the latter, which established the superiority of the computer over the human mind in computational problems, making this event a milestone in the history of artificial intelligence. Despite this, people did not lose interest in chess and instead started using computers to improve the game's comprehension further. 

Given the popularity of chess, it is not surprising that also science devoted considerable attention to this game. Early theoretical studies date back to C.~E.~Shannon \cite{shannon1950xxii}. However, only recently, thanks to the advent of the internet and online chess platforms, it has been possible to analyze a vast amount of data with the tools of statistical physics and complex systems \cite{chowdhary2022quantifying}. For instance, Refs.~\cite{blasius2009zipf, maslov2009power, perotti2013innovation} found that chess openings obey Heaps' and Zipf's laws, two statistical regularities which are often considered the footprint of complexity \cite{de2021dynamical}. Also, the history of recorded games shows long-range memory effects \cite{schaigorodsky2014memory, schaigorodsky2016study}.
Other studies focused on chess players' ratings and their evolution \cite{simkin2015chess, fenner2012discrete}, and on the popular level learning of the game \cite{ribeiro2013move}. 

One of the factors determining the complexity of chess is the vast number of different possible playable games, which C.~E.~Shannon \cite{shannon1950xxii}estimated to be around $10^{120}$. 
However, even if there are more than 71,000 different reachable positions after the first four moves, only a tiny amount is observed in real games since not all moves are equally good. The study of the sequence of initial good moves is called Chess Opening Theory. Currently, the most authoritative resource on this subject is the Encyclopedia of Chess Openings, which classifies openings in 500 different ECO codes \cite{matanovic1971}. Openings are a central part of chess, and top-level players spend a significant fraction of their time studying novel opening ideas and memorizing new lines.

Mastering Opening Theory requires a deep knowledge of chess that is typically well beyond the capabilities of amateurs. As a consequence, only world-level professional players have a complete understanding of the Opening Theory in its entirety. Here we show that by considering a large community of chess players and leveraging complex network theory, it is possible to exploit the emerging social intelligence of the community (``the wisdom of the crowd'') to overcome this limit. The idea is that even if each player in the community has a partial individual knowledge, all this knowledge can be suitably combined to obtain a complete picture of the whole Opening Theory. This approach allows us to quantify chess features that only chess experts could appreciate so far: (i) the similarity between openings; (ii) the complexity of openings; (iii) the quality of players' opening repertoires.

\section*{Results}
	\subsection*{The bipartite network of chess players and openings}
		\label{sec:bipartite}
		Network theory is one of the pillars of complexity since most complex systems spontaneously arrange into graphs, chess making no exception \cite{almeira2017structure}. Recently, bipartite networks received an increasing interest since the graph representation of many systems displays this peculiar arrangement of nodes. 
		A graph is bipartite if two classes of nodes exist such that the nodes of the same class do not connect, while links connect the nodes of different classes. 
        For example, one can represent the world trade network with the bipartite network formed by the products and by the countries exporting them.
        By leveraging that network, it is possible to obtain state-of-the-art long-term GDP forecasts \cite{tacchella2018dynamical} and to predict the industrial upgrading of countries \cite{zaccaria2014taxonomy, albora2021product}. Remarkably, we can identify bipartite networks also in chess where they can be used to gather novel insight into this game. 
		
        In technical terms, a bipartite network is a network whose nodes can be divided into two sets $\mathcal{P}$, $\mathcal{O}$ such that there are no links between nodes belonging to the same set. Denoting by $P$ the number of nodes in the first set and by $O$ the number of nodes in the second one, an unweighted directed bipartite network can be represented by a $P\times O$ matrix $\mathbf{M}$ such that $M_{po}=1$ if $p\in P$ and $o\in O$ are connected and $M_{po}=0$ otherwise. 
        In this study, we consider the bipartite network of chess players and chess openings, which we built using games played on the online chess platform \url{lichess.com} with a Blitz time control (see \blue{Data} section).
        We chose Blitz games since this format is the most played online. 
        The Lichess platform uses the \emph{Glicko-2} system to rate players (see \blue{Data} section).
        We consider the 500 chess openings with their ECO code as appearing in the ``Encyclopedia of Chess Openings'' \cite{matanovic1971,ECO}.
        Each ECO code corresponds to two nodes in the network since we distinguish between playing with White pieces or with Black pieces. For instance, if a game between player A with White and player B with Black falls under ECO code C20 (King's pawn game), then A is connected to the opening C20W (King's pawn game with White), while B to the opening C20B (King's pawn game with Black).
        In this first part of the analysis, we selected a sub-sample of chess players with a rating above $2000$ and who played at least $100$ games with Black and $100$ games with White during the period considered, from October 2015 to September 2016 (one year). This way, we ended up with a network composed of $2513$ players and $982$ openings (no player played $18$ of the 1,000 openings during the time-laps we analyzed).   
        The matrix $M$ thus satisfies
		\[
			M_{po}=
			\begin{cases}
				1 \ \text{if player}\ p \ \text{played opening}\ o\\
				0 \ \text{otherwise}
			\end{cases}
		\]
		We can then define for each player $p$ their diversification $d_p$ as the number of distinct openings they use $d_p=\sum_o^O M_{po}$ and for each opening $o$ its ubiquity $u_o$ as the number of players who used it $u_o=\sum_p^P M_{po}$.
	\subsection*{The network of chess openings}
		\label{sec:opening_network}
		\begin{figure*}[t]
		        \centering
				\includegraphics[width=0.95\textwidth]{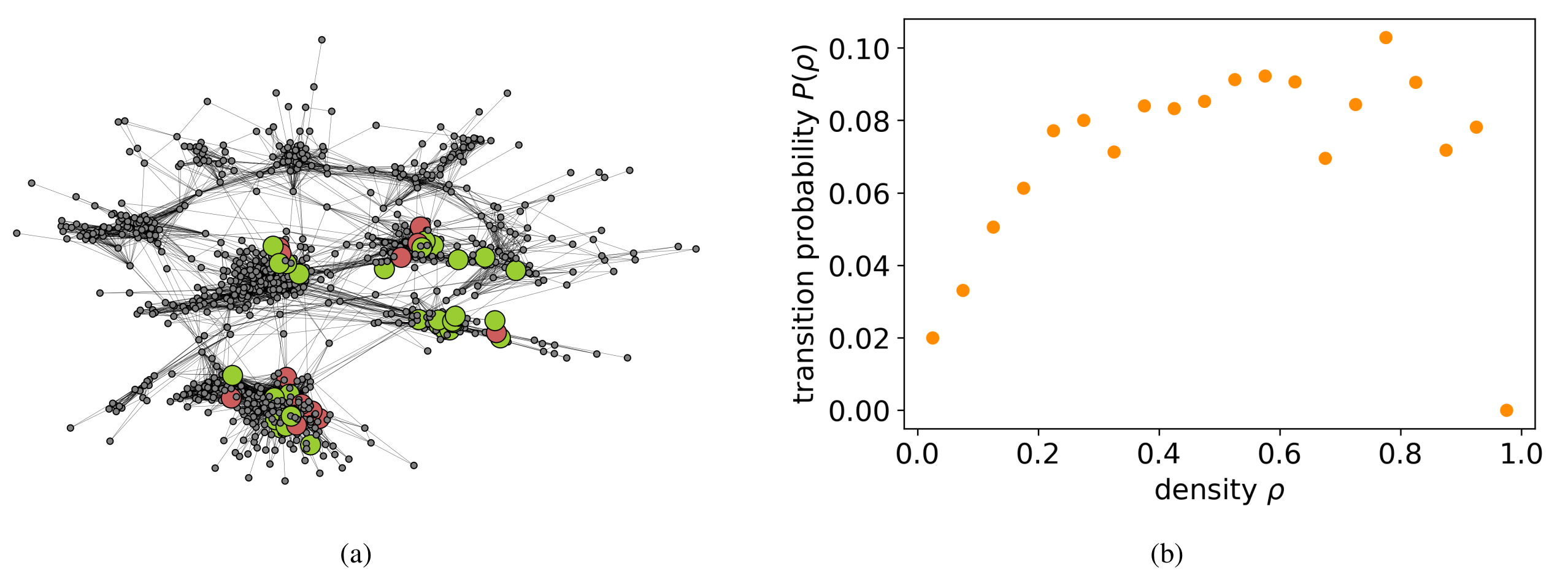}
		        \caption{\textbf{Prediction of future openings.} 
		        \textbf{a)} We plot, on the top of the opening network of Fig.~\ref{fig:network_leiden}, the openings a randomly selected player used during the period July 2016-September 2016 (green nodes) and the new openings he/she used in the period October 2016-December 2016 (red nodes). As it is possible to see, new openings are close to those openings the player already used. 
		        \textbf{b)} Probability for a never used opening to start to be played as a function of its density, defined as the fraction of neighbour openings the player already uses. This probability is increasing in density, meaning players tend to learn openings close to those they already know. The network topology defines closeness.} 
		        \label{fig:predictions}
		\end{figure*}
		Relatedness networks have been widely used in the economic complexity literature to quantify the similarity between nodes belonging to the same layer of bipartite networks \cite{hidalgo2007product}. For instance, considering the country-product network, the assumption is that two products require similar capabilities for being produced if they appear together in the export basket of many different countries or, in other words, if they co-occur often. In the same way, here we build the relatedness network of chess openings leveraging the idea that two of them are similar to play if many players play them both. We thus define the relatedness $W^*_{o_1 o_2}$ between two openings $o_1$, $o_2$ as 
		\begin{equation}
			W^*_{o_1 o_2}=\sum_p^P M_{p o_1}M_{p o_2}
			\label{eq:projection}
		\end{equation}
		However, the resulting matrix $\mathbf{W^*}$ contains many spurious co-occurrences. Two openings may often co-occur just because players with high diversification use them both by chance or because the openings are ubiquitous but not similar. In order to filter out such spurious co-occurrences, we exploit a null model, namely the Bipartite Configuration Model (BiCM) \cite{saracco2015randomizing}. The idea is to retain only those links that can not be explained only through the ubiquity and the diversification of nodes and that thus are statistically significant;  we report more details in \blue{Methods}. 
		In the following we denote by $\mathbf{W}$ the matrix resulting from this procedure, whose elements satisfy 
		\begin{equation}
			W_{o_1 o_2}=
			\begin{cases}
				1 \ \text{if}\ W^*_{o_1 o_2} \ \text{is statistically significant}\\
				0 \ \text{otherwise}
			\end{cases}
			\label{eq:validated_projection}
		\end{equation}
		and we call relatedness network of openings, or simply network of openings, the network defined by such matrix. 	
		
		We show the network of chess openings in Fig.~\ref{fig:network_leiden}. It is composed of $924$ nodes out of the initial $982$ since we filtered out isolated nodes and small components formed by pairs of nodes. We then applied the Leiden algorithm \cite{traag2019louvain} to this network to detect communities, which we indicate with different colours in the figure. There are ten clusters, three corresponding to openings played from the White perspective and the remaining seven to those played from the Black perspective. The three White clusters almost perfectly correspond to the three main choices White has as the first move: 1.~e4 ($65\%$ of games), 1.~d4 ($24\%$ of games), 1.~c4 ($3\%$ of games) (The percentages come from the Lichess database \url{https://lichess.org/analysis##0}). 
		More precisely,
		\begin{itemize}
			\item King's pawn opening (light blue), where White's first move consists in advancing the king's pawn by two squares (1. e4)
			\item Queen's pawn opening (red), characterized by White choosing as first move to advance the queen's pawn by two squares (1. d4)
			\item English opening (light green), where White opens by moving the c2 pawn by two squares (1. c4). Also Reti opening (1. Nf6) is contained in this cluster since it often involves advancing the c2 pawn by two squares as second move (2. c4)
		\end{itemize}
		Also Black clusters nicely maps to Black's main choices, but their number is higher since Black's reply also depends on White's first move. The clusters are
		\begin{itemize}
			\item e5 openings (khaki), where Black plays the move e5 as reply to a King's pawn opening (1. e4 e5) or as reply to the English opening (1. c4 e5)
			\item e6 openings (dark blue), which can be divided in two main categories, as also evident from the shape of the cluster. The French opening (1. e4 e6) on the right, where e6 comes as reply to the King's pawn opening, and Queen's pawn game openings (Catalan, Bogo-Indian, Queen's Indian and Nizmo-Indian) on the left, where Black plays e6 as second move after White having opened with her Queen's pawn (1. d4 Nf6 2. c4 e6)
			\item Sicilian defense (orange), characterized by Black replying with c5 to White's King's pawn opening (1. e4 c5). This community also contains other openings characterized by Black playing c5 such as the Symmetric English (1. c4 c5).	
			\item d5 openings (yellow), where d5 may comes as response to a Queen's pawn opening (1. d4 d5) or to a King's pawn opening as in the Caro-Kann defense (1. e4 c6 2. d4 d5)
			\item Benoni defense (Pink), which comes as response to White playing the Queen's pawn opening (1. d4 Nf6 2. c4 c5)
			\item Grunfeld and King's Indian defenses, both characterized by the same initial moves (1. d4 Nf6 2. c4 g6). It is worth remarking that both these openings are very similar to Benoni defense and this explains their closeness in the network
			\item Dutch defense, where Black responds with f5 to White's Queen's pawn opening (1. d4 f5)
		\end{itemize}
		We thus see that while clusters of White openings are determined only by White's first move, those formed by Black openings also depend on the second move Black plays. It is worth pointing out that one cannot deduce this clustered structure directly from the ECO classification.
        In fact, openings belonging to the same ECO group split into different communities, so clusters are often formed by openings belonging to different ECO groups. Chess experts do have the overall picture of this clustered structure clear in their mind.
 		
	\subsection*{Forecast of future openings}
		\label{sec:forecast}
		\begin{figure*}[t]
			\centering
			\includegraphics[width=0.7\textwidth]{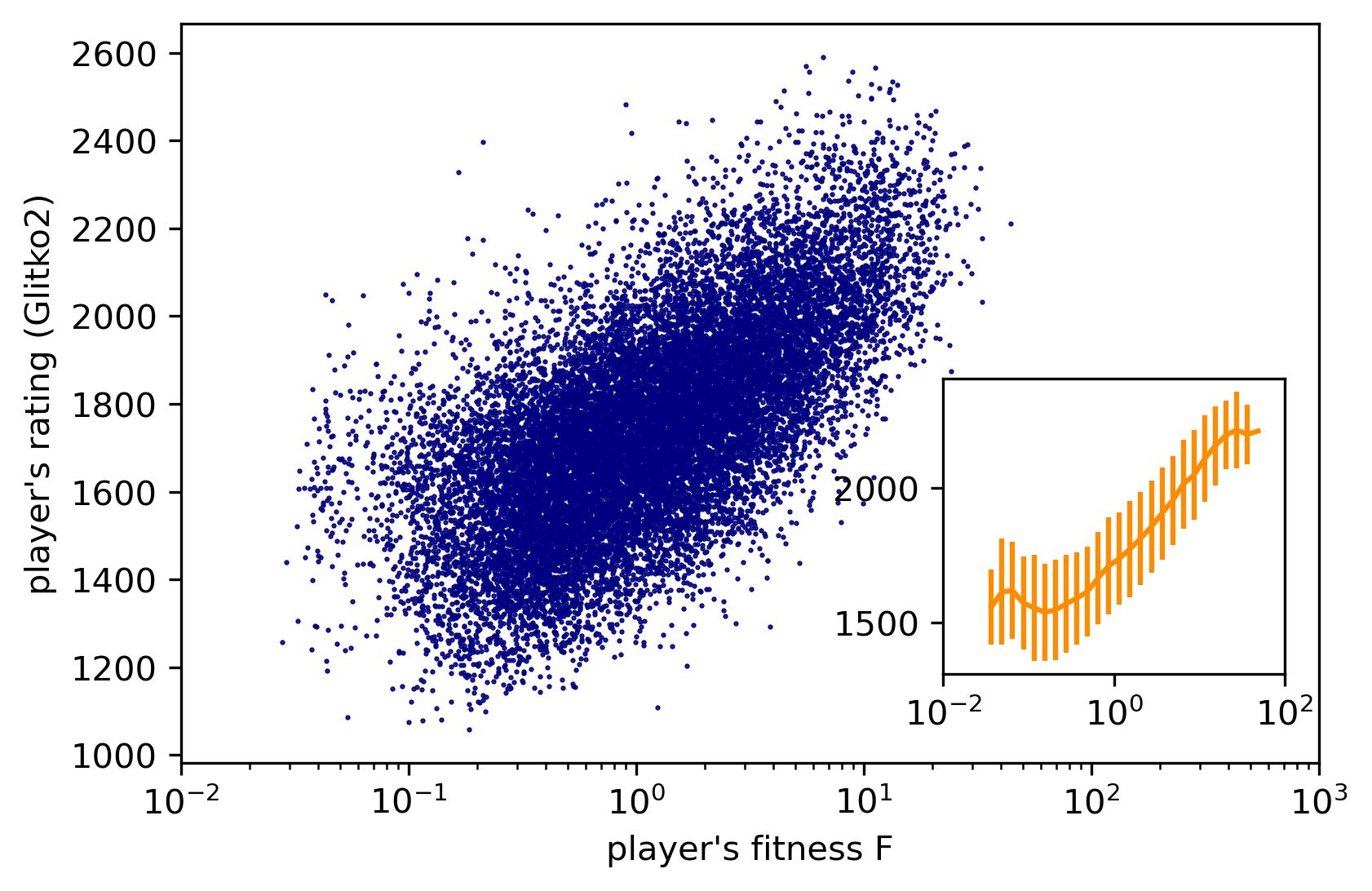}
			\caption{\textbf{Fitness of players.} Comparison between players' fitness and players' Glicko2 rating. We see a strong correlation between the two quantities, meaning that players' opening preparation is crucial in determining their ability to win games and so to reach a high rating. In the inset, we show the binned average rating as a function of the fitness. The standard deviation of the sample determines error bars.}     
			\label{fig:fitness}
		\end{figure*}
		As explained in the previous section, the relatedness between two openings quantifies how similar to play they are. Thus, the opening network allows measuring the distance between openings. We already assessed this qualitatively by inferring the clustered structure of the network. Now, we go a step further by showing that we can use the opening network to predict which openings a player will start to play in the future. The idea is that those openings close to a player's opening repertoire are easier for her to learn. We sketch this situation in Fig.~\ref{fig:predictions}a. We plotted in green the openings a randomly selected player used during July 2016-September 2016 and in red those openings she started to play during the following three months, while small grey dots are all the remaining openings the player did not use. 
		{We observe that new openings are close to those used in the first time interval, giving rise to a sort of ``adjacent possible'' effect \cite{tria2014dynamics, kauffman1996investigations}. The adoption of a brand new opening opens the possibility of playing new openings which are connected to it: the learning of Opening Theory can be seen as an expansion into the ``adjacent possible'' identified by the opening network.}                    
		
		In order to quantify the role played by the closeness of the opening network in the learning of new openings, we considered a sample of $8831$ players, forecasting the activations of new openings. We report the details about the data used in the \blue{Methods}. We denote by $M_{po}^{i}$ the adjacency matrix obtained considering the matches played by these players in the period July 2016-September 2016 and by $M_{po}^{f}$ the matrix corresponding to October 2016-December 2016. In these terms, the activations $a$ are those openings not played in the first period, that is $M_{pa}^{i}=0$. We then define the density $\rho_{pa}$ of activation $a$ for player $p$ as \cite{hidalgo2007product}
		\[
			\rho_{pa} = \frac{\sum_o W_{ao}M_{po}^{(i)}}{\sum_o W_{ao}},
		\]
		where $\mathbf{W}$ is the adjacency matrix of the openings network previously defined. The density $\rho_{pa}$ is thus the fraction of openings connected to $a$ the player $p$ already used in the first time period. We define the transition probability $P(\rho)$ as the probability an activation with density $\rho$ is played in the second period considered, i.e.\ 
		\[
			P(\rho) = \frac{\sum_p\sum_o \ton*{1-M_{po}^{(i)}}M_{po}^{(f)}\delta(\rho-\rho_{po})}{\sum_p\sum_o \ton*{1-M_{po}^{(i)}}\delta(\rho-\rho_{po})}.
		\]
		This quantity is plotted as function of the density in Fig.~\ref{fig:predictions}b. We see that the transition probability is an increasing function of the density and for large values of $\rho$ the probability for an opening to start to be played is about four times larger than at $\rho\approx 0$. This plot indicates that the density can be used to forecast the adoption of new openings. Therefore, we define the transition predictor $y^{\mathrm{pred}}_{pa}$ as
		\begin{equation}
			\begin{cases}
				y^{\mathrm{pred}}_{pa} = 1 \ \text{if} \ \rho_{pa}\geq \beta\\
				y^{\mathrm{pred}}_{pa} = 0 \ \text{otherwise}
			\end{cases}
			\label{eq:predictor}
		\end{equation}
		where $\beta$ is the density threshold separating openings that are predicted to be used from those that are predicted to remain unused. We tested the performance of this predictor on the activation using as ground truth the bipartite matrix of the second period $y^{\mathrm{true}}_{pa}=M_{pa}^{(f)}$. Its Best F1 Score, corresponding to $\beta=0.2$, is $0.16$, to be compared to Best F1 Score$=0.04$ obtained with a random predictor. We report The definition of the Best F1 Score and more details about the predictor's performance in the \blue{Methods}. This score, even if not notably high, is a good result for several reasons. First, we notice that players usually have many possible openings to start to play, but they use only a few. Second, in the similar context of the bipartite network country-product, state-of-the-art machine learning techniques reach a Best F1 Score$\approx 0.04$ \cite{albora2021product}. Finally, here we are not interested in obtaining the best forecast possible. Instead, we want to demonstrate that our opening network contains useful information.
		
	\subsection*{The fitness of players and the complexity of openings}
		\label{sec:fitness_complexity}
		\begin{figure*}[t]
			\centering
			\includegraphics[width=0.85\textwidth]{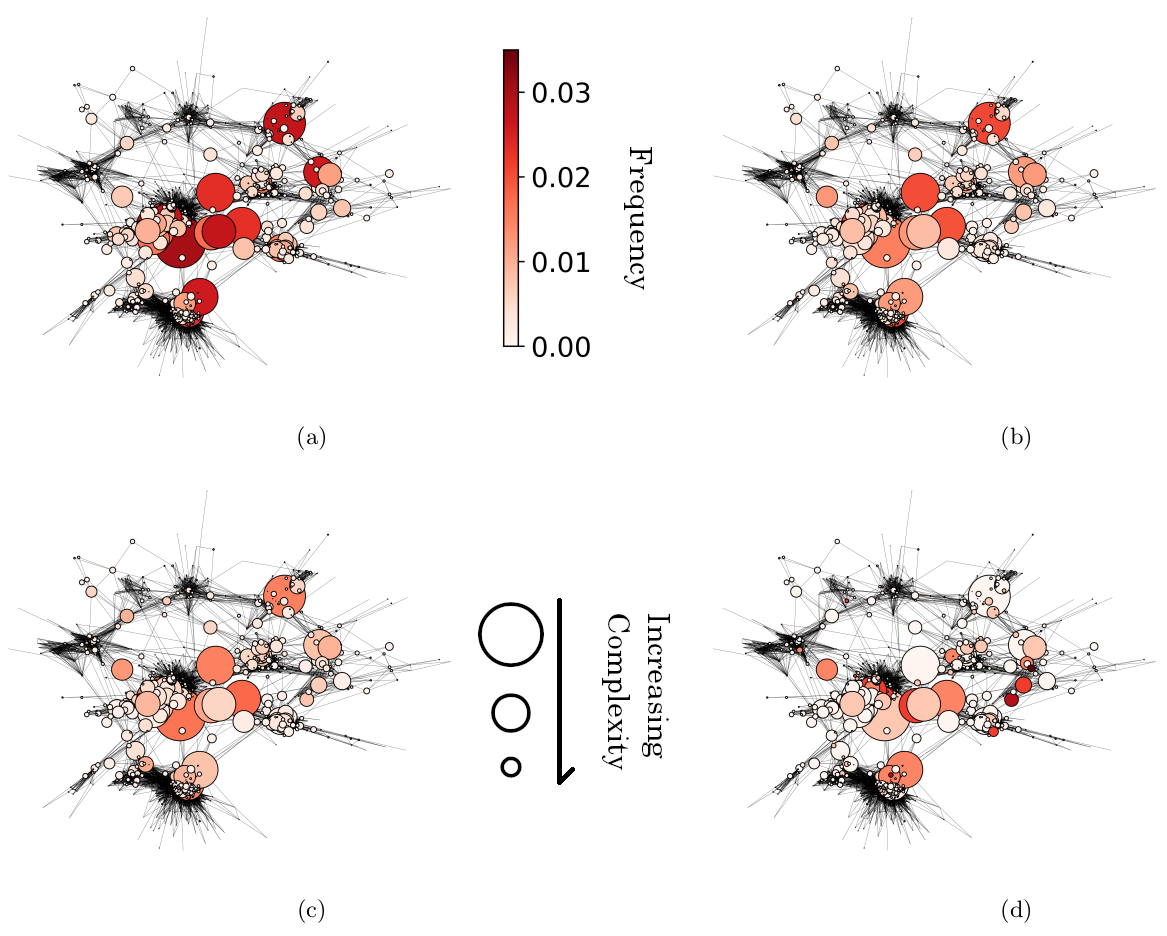}
		 	\caption{\textbf{Complexity of openings.} \textbf{a)} Average opening repertoire of low-rated chess players visualized using the opening network. We selected all players with ratings between $1500$ and $1600$, and for each opening, we computed how frequently they use it. Dark colours correspond to frequently used openings, while light colours to infrequently used ones. The size of nodes is inversely proportional to the openings' complexity, so small nodes are hard-to-play openings, while big nodes are easy-to-play openings. We see that low-rated players tend to use low-complexity openings frequently. \textbf{b)} As panel a, but for players with ratings between $1900$ and $2000$. \textbf{c)} As panel a, but for players with ratings between $2300$ and $2400$. \textbf{d)} As panel a, but considering the opening repertoire of world champion Magnus Carlsen. As a player's rating increases, low complexity openings tend to be used less frequently, while the frequency of more complex openings increases.}
		    \label{fig:complexity_openings}
		\end{figure*}
		Not all openings are equally easy to play since some require a deep knowledge of chess theory. 
		This is an aspect our opening network does not capture, but, as we will show, we can exploit the information contained in the player-opening bipartite network to estimate how difficult to play an opening is.
		First of all, we introduce the normalized matrix $\mathbf{N}$ whose entries $N_{op}$ are given by the fraction of games in which player $p$ used opening $o$
		\[
			N_{op}=\frac{n_{op}}{\sum_{o}n_{op}},
		\]
		where $n_{op}$ is the number of times player $p$ chose opening $o$. 
		This matrix defines a bipartite network of players and openings with links weighted by the frequencies of the played openings. 
		We then exploit the Economic Fitness and Complexity (EFC) algorithm \cite{tacchella2012new} to compute the complexity of openings $Q_o$ and the fitness of player $F_p$. The former quantifies how tough to play openings are, while the latter measures the opening skills of players. The EFC algorithm is a recursive non-linear map which has been successfully applied to rank nodes of bipartite networks \cite{tacchella2012new, sbardella2018green, cimini2014scientific, dominguez2015ranking} and which, in its original form, is defined by the following map
		\begin{equation}
	        \begin{cases}
				Q_o^{(t+1)}=\frac{1}{\sum_p N_{po}\frac{1}{F_p^{(t)}}}\\
				F_p^{(t+1)}=\sum_o N_{po}Q_o^{(t)}
			\end{cases}\quad\quad
	        \begin{cases}
	            Q_o^{(t)} = \frac{\tilde{Q}_o^{(t)}}{\mean*{\tilde{Q}_o^{(t)}}_o}\\
	            F_p^{(t)} = \frac{\tilde{F}_p^{(t)}}{\mean*{\tilde{F}_p^{(t)}}_p}
	        \end{cases}
			\label{eq:fitness_complexity}
		\end{equation}
		where by $t$ we denote the iteration step, while $Q_o$ and $F_p$ are given by the fixed points of such a map $Q_o=\lim_{t\to\infty}Q_o^{(t)}$ and $F_p=\lim_{t\to\infty}F_p^{(t)}$.  The original map of Eq.~(\ref{eq:fitness_complexity}) suffers of convergence issues so that we use a slightly different map, the non-homogeneous EFC (NHEFC), which delivers almost the same results but with no convergence problems \cite{servedio2018new}. We report the details of the NHEFC algorithm, its convergence and its implementation in the \blue{Methods}. We add here few considerations about Eqs.~\eqref{eq:fitness_complexity}. The first expression implies that an opening has low complexity if low-fitness players play it. This is no surprise since one expects low-fitness players to use only simple-to-play openings. The second expression states that the weighted average of the complexity of openings a player uses is the player's fitness, whereas the frequencies of the played openings determine the weights.
        Note that this differs from the standard EFC algorithm. In the original formulation, the fitness at the first iteration is given by the diversification $F_p^{(1)}=d_p$, while in our implementation $F_p^{(1)}=1$. 
		
		As already mentioned, openings play a significant role in chess games, so we expect players with high fitness also to have a high rating. In order to assess if this is the case, we considered a sample of $18,253$ players. We built the corresponding frequency matrix $\mathbf{N}$ and we applied the NHEFC algorithm. Again, details on the data are available in the \blue{Methods}. In Fig.~\ref{fig:fitness}, we show the scatter plot with players' rating against their fitness. There is a strong correlation between these two quantities, as confirmed by a Spearman correlation coefficient of $0.64$. In the inset, we report the average rating as a function of the fitness (error bars defined by the standard deviation). Remarkably there are two flat regions corresponding to low-rated and high-rated players, respectively, in which an increase in the fitness does not affect the rating. This means that beginners should first learn the basic ideas of chess before focusing on opening theory. At the same time, for high-rated players, it is not very easy to improve only by focusing on openings since other aspects, such as endgames or time management, also start to be very relevant. 
		
		Finally, we studied the complexity's meaningfulness by analyzing how players' opening repertoire changes depending on their rating. This is done in Figs.~\ref{fig:complexity_openings}a, \ref{fig:complexity_openings}b and\ref{fig:complexity_openings}c, where we considered players with rating in the ranges $1500-1600$, $1900-2000$ and $2300-2400$ and we plotted their average opening repertoire on the opening network. The sizes of nodes are inversely proportional to their complexity so that easy-to-play openings are represented as large circles, while dark colours indicate frequently used openings. We see that the opening repertoire of low-rated players is concentrated mainly on low-complexity openings, which are less frequently used by high-rated players. Analogously, Fig.~\ref{fig:complexity_openings}d shows the opening repertoire of world champion Magnus Carlsen (nickname DrNykterstein): here, some of the less complex openings are completely absent, while several small nodes, corresponding to more complex openings, are dark and so frequently used. This confirms that complexity is a good indicator to quantify the difficulty of openings.
		
\section*{Discussion}
	Chess is probably the most fascinating board game, and, despite its old origins,  a considerable number of people still play it all around the world. Opening Theory is one of the most complex aspects of this game and requires years of study and practice to be mastered. As a consequence, amateurs only have minimal knowledge of chess openings. However, as we showed in this work, it is possible to extract information about the Opening Theory that goes well beyond the knowledge of single players by considering a whole online chess community. This allows to analyze aspects and answer questions that otherwise would require the help of chess experts. As a first step, we use the player-opening bipartite network to obtain the one-layer relatedness network of chess openings. Two openings linked in this network are similar to play; thus, the opening network quantifies the distance among chess openings. We obtain ten clusters by applying a community detection algorithm to such a network. Three of them contain openings seen from White's perspective and almost perfectly correspond to White's three main choices for her first move (1. e4, 1. d4. 1.c4). The remaining seven correspond to Black's most common reply to White's first move. This structure is non-trivial and can not be directly derived from the Encyclopedia of Chess Openings (ECO) classification, thus showing that our entirely data-driven network-based approach allows unveiling hidden similarities between chess openings. We then exploit the opening network to forecast which openings a player will start to play in future. We do this relying on the assumption that those openings that are ``surrounded'' by openings the player already uses should be easy to learn since they are similar to what she already knows. In practical terms, we introduce density, which measures the surroundedness of openings and we show that the probability for an opening to be used is an increasing function of this quantity. Forecasts based on the density reach a best $F1$ score about four times larger than those obtained by a random predictor. Finally, we exploit a variant of the Economic Fitness and Complexity algorithm to obtain a data-driven definition of openings' complexity and players' fitness. The latter quantifies how skilled players are in openings and shows a $0.64$ correlation with the rating of players. The former measures how challenging to play openings are and we demonstrate its meaningfulness by showing that low-rated players tend to focus on low-complexity openings, while high-rated players also exploit complex openings.
	
	We conclude by pointing out that the analysis here discussed also opens the possibility of devising personalized recommendations for chess players. For instance, by leveraging the opening network, it is possible to suggest to players openings they could learn easily and, taking into account their fitness and the complexity of openings, such suggestions can be modulated based on players' skill. Moreover, the opening network combined with the complexity allows visualizing in a single image the opening repertoire of a player, thus making it possible to understand weaknesses in her opening preparation or compare two players effortlessly. This makes us think these tools can be helpful to scientists interested in studying the game of chess and to any chess player willing to improve their opening repertoire.

\section*{Methods}
	\subsection*{Data}
		We used data gathered from the online chess platform \url{lichess.com} for carrying out our analysis. These data are freely available at \url{https://database.lichess.org/}. 
		\subsubsection*{Definition of Blitz games}
		We only selected Blitz games since this format is the most played online. 
		Two integer numbers define the time control of one player, e.g., ``X+Y'', where X is the clock initial time in minutes and Y is the clock increment in seconds. 
        Lichess considers a game to fall in the Blitz category if the estimated time of an average match, which is supposed to end in 40 moves per player, $T=X\times60+40\times Y$ in seconds per player is such that $179\le T < 479$.
        For instance, the estimated duration of a ``5+4'' game is $5 \times 60 + 40 \times 4 = 460$ seconds for each player, so that ``5+4'' games belong to the Blitz category.
        \subsubsection*{Players' strength}
        Lichess estimates the ability of players with the \emph{Glicko-2} rating system \cite{glicko2}.
        After each rated match, the Glicko-2 value of both players changes according to the result of the match. Generally, who win get their Glicko-2 value increased. Therefore, the higher the Glicko-2 value, the more skilled are the players. 
        Many gaming platforms use the Glicko-2 system. With respect to the ELO system adopted by the International Chess Federation (FIDE -- from the French translation --), Glicko-2 takes into account confidence intervals, i.e., an uncertainty in the assigned rating.
        \subsubsection*{List of chess openings}
        We collect the list of 500 chess openings with their ECO code according to the ``Encyclopaedia Of Chess Openings'' \cite{matanovic1971,ECO}. A comprehensive list of openings can be extracted from \url{https://en.wikipedia.org/wiki/List_of_chess_openings}. We consider each opening from the point of view of both the White player and Black player, so that, in fact, the number of total openings in our analysis is 1,000.
		\subsubsection*{Bipartite network and relatedness network}
			In order to build the bipartite network we used blitz games played from October 2015 to September 2016 and we applied the following filtering procedure:
			\begin{itemize}
				\item we selected only players with rating above $2000$ because we expect low rated player to use some openings just by chance; considering also them would add noise to the projected opening network;
				\item we selected the games where the rating of the two players differs at most by $50$, since if the difference between the two players is very high, then the high rated player could use a bad opening just to have more fun;
				\item we removed players playing less than $100$ games with White and $100$ games with Black so to have for each player a large statistic.
			\end{itemize}
			We ended up with a total of $472,183$ games involving $2,513$ players and $982$ different openings. 
		\subsubsection*{Forecast}
			In order to forecast the use of openings we used two different time periods. We used blitz game data from July 2016 to September 2016 to compute the density, while the goodness of predictions have been assessed using data ranging from October 2016 to December 2016. Note that this last period does not overlap with that considered in the building of the opening network, this being an important point in order to reliably assess the goodness of the predictions. Also in this case, we retained only games with a maximum rating difference of $50$ and we selected only players who did $50$ matches with the White pieces and $50$ games with the Black ones, so to have a large enough statistic for computing the density, ending up with a total of $88,31$ players.
		\subsubsection*{Fitness and Complexity}
			The fitness of players and the complexity of openings has been obtained using games played in the period October 2015 to September 2016. Differently from what done for building the bipartite network, we considered all ratings, but we made the same filtering with respect to the rating difference and the number of games played. This gives us a total of $3,746,135$ games played between $18,253$ players who used $988$ different openings. We also used the Lichess Elite Database \url{https://database.nikonoel.fr/} to get $138$ games played by Magnus Carlsen in November 2021. 
	\subsection*{Validation of projected networks}
		In order to build the relatedness network of chess openings we project the bipartite matrix $M_{po}$ connecting players to openings. As explained in the main text, this can be done by using Eq.~\eqref{eq:projection}, that is 
	    \[
	        W^*_{o_1 o_2}=\sum_p^P M_{p o_1}M_{p o_2}.
	    \]
	    In this way, we obtain the matrix $\mathbf{W^*}$ connecting those openings appearing together in the opening repertoires of many different players. However, such a matrix is generally almost fully connected due to spurious co-occurrences. Consequently, one has to use a null model to retain only statistically significant links and filter out the spurious ones. In this work, we exploit the Bipartite Configuration Model (BiCM) \cite{saracco2015randomizing, saracco2017inferring, vallarano2021fast}, which is based on the theory of exponential random graph; in particular we used the python bicm library \url{https://bipartite-configuration-model.readthedocs.io/en/latest/}. The BiCM is based on a canonical ensemble of random graphs defined by constraining (on average) the degree sequences of both node sets (so ubiquity and diversification). We then obtain the probability distribution of such an ensemble by maximizing Shannon entropy under this constraint. This probability distribution reads 
	    \[
	        P(\bar{\mathbf{M}}|\{\theta_p\}, \{\mu_o\}) = \frac{\me ^{-H(\bar{\mathbf{M}}|\{\theta_p\}, \{\mu_o\})}}{Z(\{\theta_p\}, \{\mu_o\})},
	    \]
	    where
	    \begin{itemize}
	        \item $\bar{\mathbf{M}}$ is the adjacency matrix of the random bipartite network
	        \item $\{\theta_p\}$ and $\{\mu_o\}$ are the Lagrange multipliers associate respectively to the diversification of players $\{d_p\}$ and to the ubiquity of openings $\{u_o\}$
	        \item $H(\bar{\mathbf{M}}|\{\theta_p\}, \{\mu_o\})$ is the Hamiltonian, defined as 
	        \[
	        H(\bar{\mathbf{M}}|\{\theta_p\}, \{\mu_o\})= \sum_p \theta_p d_p(\bar{\mathbf{M}}) + \sum_o \mu_o u_o(\bar{\mathbf{M}}).
	        \]
	        \item $Z(\{\theta_p\}, \{\mu_o\})$ is the partition function of the Hamiltonian
	        \[
	            Z(\{\theta_p\}, \{\mu_o\})=\sum_{\bar{\mathbf{M}}}\me ^{-H(\bar{\mathbf{M}}|\{\theta_p\}, \{\mu_o\})}.
	        \]
	    \end{itemize}
	    It can be shown that the probability distribution factorizes and the probability that for player $p$ and opening $o$ to be connected in the random network is
	    \[
	        p_{po} = \frac{1}{\me^{\theta_p+\mu_o}+1},
	    \]
	    while the numerical values of the Lagrange multipliers are obtained by solving the system 
	    \[
	        \begin{cases}
	            d_p = \sum_s p_{po}\\
	            u_o = \sum_j p_{po}.
	        \end{cases}
	    \]
	    
	    Once we have the probability of the links we can generate $N$ random bipartite networks and, for each of them, compute the projected matrix $\bar{\mathbf{W^*}}$. At this point, we validate the links of $\mathbf{W^*}$, setting to one all those links which are in $99\%$ of the cases larger than the corresponding link in the random matrices. We set to zero the rest of the links. In this way,  we obtain the validated matrix $\mathbf{W}$ defined in Eq.~\eqref{eq:validated_projection}. Here the threshold $99\%$ is arbitrary and sets the confidence level.
	\subsection*{Goodness of prediction test}
		Using the density measure we introduced above, it is possible to predict if a player will start to use a certain opening. Here we discuss about how to evaluate the goodness of these predictions. We recall that our predictions, denoted by $y^{\mathrm{pred}}_{pa}$, are defined by Eq.~\eqref{eq:predictor}
		\begin{equation*}
			\begin{cases}
				y^{\mathrm{pred}}_{pa} = 1 \ \text{if} \ \rho_{pa}\geq \beta\\
				y^{\mathrm{pred}}_{pa} = 0 \ \text{otherwise}
			\end{cases}
		\end{equation*}
		and are obtained using data in the period July 2016-September 2016, while the ground truth is obtained from data in the period October 2016-December 2016 and is defined as $y^{\mathrm{true}}_{pa}=M_{pa}^{(f)}$. We recall that by $a$ we denote the activations, so those openings not used by player $p$ during the first period; $\beta$ is the density threshold separating the openings that we predict will be played from those that we predict will not. 
        The most common metrics to evaluate the goodness of predictions are \cite{powers2020evaluation}:
		\begin{itemize}
			\item \textbf{Precision}, defined as the ratio between true positives and positives (true positives plus false positives). In our case is the ratio between the number of activations we correctly predict to be played in the second period and the total number of activations we predict to be started to play. High Precision means that openings that are predicted to be played are often played in the second period;
			 \item \textbf{Recall}, given by the ratio of true positives and the sum of true positives and false negatives. A high recall implies that openings that are predicted not to be played are rarely played in the second period;
			 \item \textbf{F1 Score}, defined as the harmonic mean of Precision and Recall. The F1 score is particularly suitable when the data are unbalanced, meaning that there are many more negatives than positives (or vice versa). A high F1 Score thus implies that both the Precision and the Recall are also high. 
		\end{itemize}
		All these indicators depend on the threshold $\beta$. We then compute the \textbf{Best F1 Score} using the threshold $\beta$ which maximizes the F1 Score, that is 
			 \[
			 	\text{Best F1 Score}=\max_{\beta}\qua*{\text{F1 Score}(\beta)}.
			 \]
			 In the case under consideration the best threshold is $\beta=0.2$. Using this threshold we obtain 
			 \begin{itemize}
			 	\item Precision $0.10$
			 	\item Recall $0.47$
			 	\item Best F1 Score $0.16$
			 \end{itemize}
	\subsection*{The Economic Fitness and Complexity algorithm}
		The Economic Fitness and Complexity (EFC) algorithm \cite{tacchella2012new, cristelli2013measuring} is an iterative non-linear map initially designed to study the Country-Product bipartite network. It allows to compute the fitness, which is an indicator of the manufacturing capabilities of a country, and the complexity, which quantifies how sophisticated and challenging it is to produce a good. This approach outperforms other techniques and allows to obtain state-of-the-art long-term GDP forecasts \cite{tacchella2018dynamical}. In our study, we apply this algorithm to the player-opening bipartite network, thus associating to each player $p$ a fitness $F_p$. The higher the fitness, the more challenging openings the player plays. To each opening $p$ we associate a complexity $Q_p$, which measures how difficult is that opening to play. These quantities, as mentioned above, are defined through a non-linear map given by Eqs.~\eqref{eq:fitness_complexity}.
	    The iteration of Eqs.~\eqref{eq:fitness_complexity} leads to a fixed point which has been proved to be stable and non-dependent on initial conditions \cite{pugliese2016convergence}. However, the EFC algorithm as defined above, in some situations has convergence issues, so we decided to follow the Servedio et al.~approach \cite{servedio2018new} to estimate players' fitness and the complexity of openings. Instead of Eqs.~\eqref{eq:fitness_complexity} we use the non-homogeneous map
	   	\[
	        \begin{cases}
				P_o^{(t+1)}=1+\sum_p \frac{N_{po}}{F_p^{(t)}}\\
				F_p^{(t+1)}=\delta^2+\sum_o \frac{N_{po}}{P_o^{(t)}}
			\end{cases}
		\]
		where $\delta$ is a parameter that can be taken arbitrarily small and does not influence the fixed point of the map; here we use $\delta = 10^{-3}$. We denote by $P_o$ and $F_p$ the fixed point of the map, in these terms the complexity of openings $Q_o$ is recovered as 
		\[
			Q_o = \frac{1}{P_o-1}
		\]	
		while the fitness of players are simply given by $F_p$. 
%


\section*{Acknowledgments}

VDPS acknowledges the support of the Austrian Research Promotion Agency FFG under grant 882184. GDM acknowledges the project ``Social and Economic Complexity'' of Enrico Fermi Research Center.
The authors of this paper are amateur chess players with no experience in official tournaments. Therefore, to validate the results from a chess theory point of view, they approached a renowned chess grandmaster who suggested some critical corrections to the paper. They offered him/her the deserved co-authorship of this paper. Unfortunately, the GM asked to remain anonymous, and they respect his/her decision. They are very grateful for the detailed suggestions provided.

\section*{Author contributions statement}

GdM and VDPS designed the study; GdM wrangled, analyzed the data and carried out most of the study; all of the authors contributed equally to the writing of the document.

\section*{Additional information}
\textbf{Competing interests}\\
The authors declare no conflict of interest.


\end{document}